\RequirePackage{fix-cm}
\documentclass[smallextended]{svjour3}       
\smartqed  
\usepackage{graphicx}

\usepackage{natbib}
\usepackage{url} 
\usepackage{amsmath}

\newcommand{\ds} {\displaystyle}
\newcommand*\dif{\mathop{}\!\mathrm{d}}

\begin{document}

\title{Clustering Via Finite Nonparametric ICA Mixture Models\thanks{%
The authors gratefully acknowledge the support of Award No.~DMS--1209007 from the National Science Foundation.}
}


\author{Xiaotian Zhu         \and
        David R.~Hunter 
}


\institute{X. Zhu \at
              Natera Inc., San Carlos, CA \\
           \and
           D.~R.~Hunter (corresponding author) \at
              Department of Statistics, Pennsylvania State University, University Park, PA \\
              Tel.: +1 814-865-1348\\
              Fax: +1 814-863-7114\\
              \email{dhunter@stat.psu.edu}           
}

\date{Received: date / Accepted: date}

\maketitle

\begin{abstract}
We propose a novel extension of nonparametric multivariate finite mixture models by dropping the 
standard conditional independence assumption and incorporating the independent component analysis 
(ICA) structure instead.  This innovation extends nonparametric mixture model estimation methods
to situations in which conditional independence, a necessary assumption 
for the unique identifiability of the parameters in such models, is clearly violated.
We formulate an objective function in terms of penalized smoothed Kullback-Leibler distance and 
introduce the nonlinear smoothed majorization-minimization independent component analysis (NSMM-
ICA) algorithm for optimizing this function and estimating the model parameters. 
Our algorithm does not require any labeled observations {\em a priori}; it may be used for fully
unsupervised clustering problems in a multivariate setting.
We have implemented 
a practical version of this algorithm, which utilizes the FastICA algorithm, in the R package icamix. 
We illustrate this new methodology using several applications in unsupervised learning and image processing.
\end{abstract}



\section{Introduction}
\label{sec:introduction}

Cluster analysis, or clustering,  is one of several 
general approaches to the problem of unsupervised learning, that is, classification when no class labels are given.
In practice, clustering is often based on heuristic ideas and intuitive measures,
such as hierarchical clustering or k-means clustering, that do not assume a probability model. 
By contrast, this paper focuses on 
model-based clustering, in which
data are viewed as coming from a 
mixture of probability distributions, each representing a cluster.  
Typically, the distributions are assumed to come from a parametric family such as normal
\citep{fraley1998many,banfield1993model,fraley2002model}, and 
group membership is learned from data by estimation algorithms that are often variations 
of the expectation-maximization method described by \cite{dempster1977maximum}. 
Since the early work of 
\cite{wolfe1963object} and others, the literature on model-based clustering has expanded enormously.
Indeed, 
there are several book-length treatments of mixture models, such as \cite{Fruhwirth2006} and \cite{mclachlan2004}.

The advent of easily accessible computing power has given rise to semi- and non-parametric 
methods that avoid the standard assumption that the
cluster densities come from a known parametric family, and applications and extensions 
of these methods are growing more common in the literature.
A semiparametric model-based clustering analysis for 
DNA microarray data can be found in \cite{han2006semi}. \cite{azzalini2007clustering} propose nonparametric 
density estimation using Delaunay triangulation for clustering via identification of subpopulations with 
regions with high density of the underlying probability distribution.  \cite{li2007nonparametric} develop a  
clustering approach based on mode identification by applying new optimization techniques to a nonparametric density estimator. 
\cite{vichi2008fitting} fits semiparametric clustering models to dissimilarity data. 
In \cite{zhang2009semiparametric}, a semiparametric model is introduced to account for varying impacts of factors over 
clusters by using cluster-level covariates. 
\cite{mallapragada2010non} propose a non-parametric mixture model (NMM) for data clustering. 
\cite{guglielmi2014semiparametric} fit Bayesian semiparametric logit models to grouped data of in-hospital survival 
outcomes of patients hospitalized with ST-segment Elevation Myocardial Infarction diagnosis. 
Certain mixtures of linear regressions also fall under the category of semiparametric model-based clustering.  
For instance, \cite{hunter2012semiparametric} present an algorithm for estimating parameters in a mixture-of-regressions 
model in which the errors are assumed to be independent and identically distributed but no other distributional assumption is made. 
\cite{huang2013nonparametric} propose nonparametric finite mixture-of-regression models for analysis of U.S.\ housing 
price index (HPI) data. 
\cite{vandekerkhove2013estimation} studies estimation of a semiparametric mixture-of-regressions model of two 
components when one component is known. 
\cite{bajari2011note} views a game abstractly as a semiparametric mixture distribution and studies 
the semiparametric efficiency bound of this model. 
Finally, \cite{butucea2014semiparametric} consider a semiparametric mixture of two distributions 
that are equal up to a shift parameter. 

The current article combines recent advances in methods for 
multivariate non-parametric finite mixture models 
under an assumption that we refer to as the 
{\em conditional independence assumption} with another method, called independent components
analysis (ICA), that solves one of the main drawbacks of this assumption.  
To illustrate this drawback, let us first introduce the modeling framework:
Suppose that $r$-dimensional vectors ${\bf{Y}}_i = ({{Y}}_{i1}, {{Y}}_{i2}, ..., {{Y}}_{ir})^\top$, $1\leq i\leq n$, are a simple 
random sample from a finite mixture of $m>1$ components with positive mixing proportions 
$\lambda_1, \lambda_2, ..., \lambda_m$ that sum up to 1, and (Lebesgue measurable) density functions 
$q_1, q_2, ..., q_m$ respectively.  This gives the mixture density
\begin{equation} \label{MixtureDensity}
d({\bf y})=\sum\limits_{j=1}^{m}\lambda_jq_{j}({\bf y})
\end{equation}
for ${\bf y}\in R^r$.  The sole assumption imposed on the densities 
$q_1, q_2, ..., q_m$ is that
the coordinates of ${\bf y}$ are independent
given the component from which ${\bf y}$ is sampled, so that
Equation~(\ref{MixtureDensity}) becomes 
\begin{equation}\label{condInd}
d({\bf y})
=\sum\limits_{j=1}^{m}\lambda_j\prod\limits_{k=1}^{r}q_{jk}(y_k).
\end{equation}
The basic idea of conditional independence is outlined by \cite{hall2003nonparametric}, 
and, notably,
\cite{allman2009identifiability} prove the generic 
identifiability of the parameters in Equation~(\ref{condInd}) for $r\geq 3$
under some weak assumptions.
\cite{chauveau2015semi} present a survey of the
growing literature on the theory and algorithmic treatment of model~(\ref{MixtureDensity}) under the
conditional independence assumption.  
The algorithms in this article have their roots in the EM-like algorithm of 
\cite{benaglia2009like}, which is later modified by
\cite{levine2011maximum}.
An alternative estimation method to EM-like algorithms is the method of moments 
approach described in \cite{anandkumar2012}.
More recently, \cite{bonhomme2016a} and \cite{bonhomme2016b}
provide new estimation algorithms for this model and prove consistency and asymptotic normality
of the estimators, an important statistical innovation missing from earlier work.
Related work on mixtures of nonparametric hidden Markov models (HMMs) is presented by
\cite{gassiat2016a}, \cite{gassiat2016b}, and \cite{decastro2016}; these authors describe links between
their HMM work and the model of Equation~(\ref{condInd}).

\begin{figure}[htb]
    \centering
    \includegraphics[width=.45\textwidth]{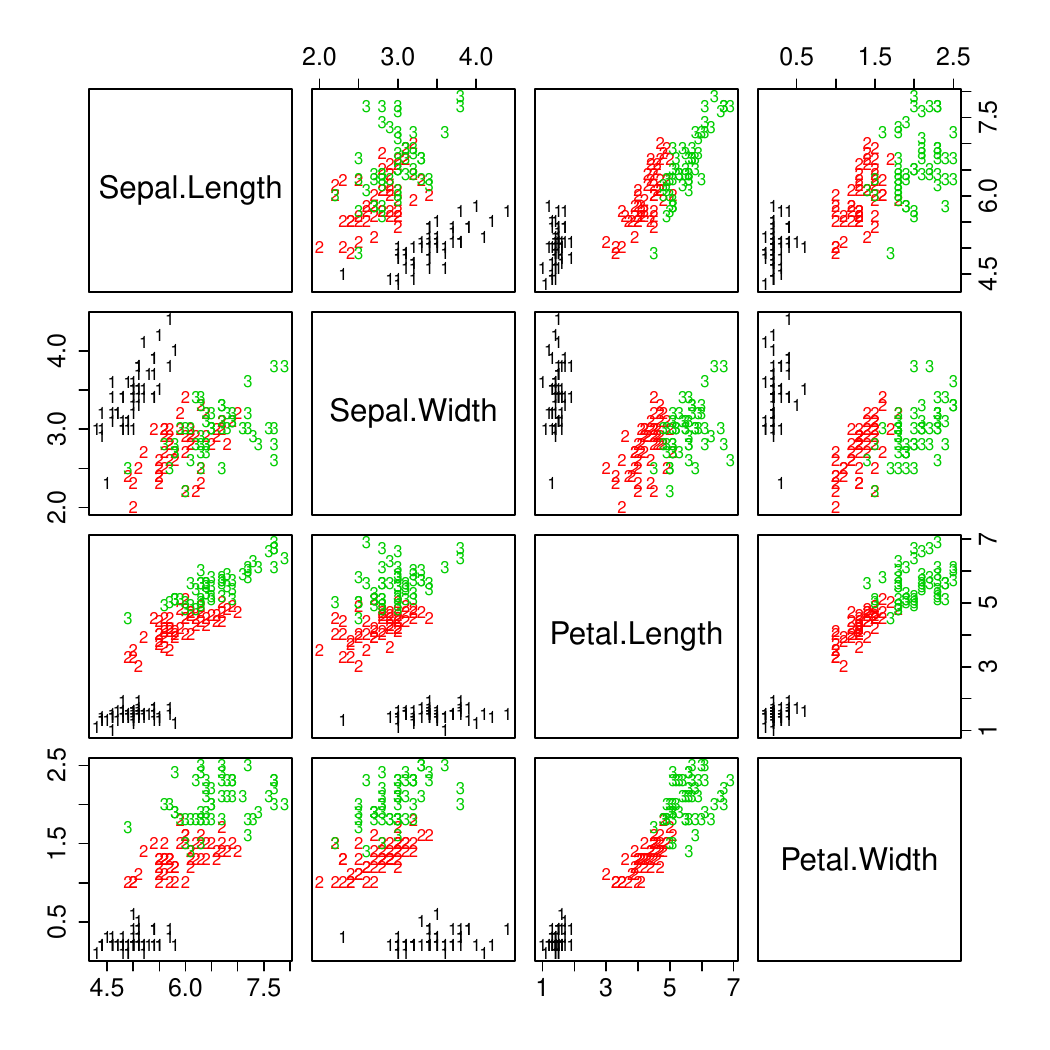}
    \includegraphics[width=.45\textwidth]{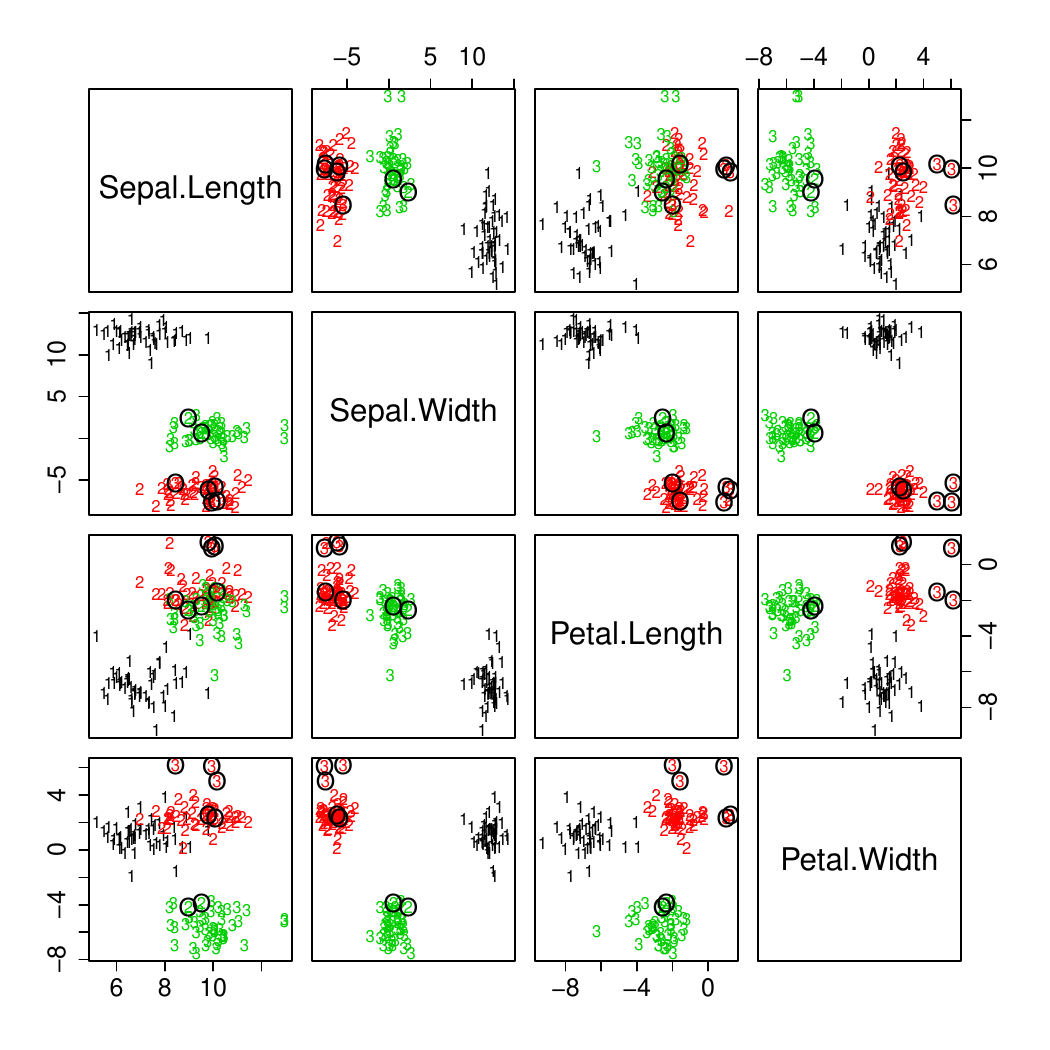}
    \caption{(a) At left, the four-dimensional iris dataset with $n=150$ and three distinct species.
    (b) At right, the same dataset classified according to our algorithm into three groups, 
    where each group is transformed linearly in an attempt to achieve conditional independence.
    The numbers are the correct species labels, and the seven points misclassified by the algorithm
    are circled.}
    \label{fig:iris}
\end{figure}

Although the conditional independence assumption of Equation~(\ref{condInd}) is
important theoretically due to its guarantee of identifiable parameters despite essentially no assumptions other than
$r\ge3$, it is clearly not appropriate for some clustering problems.
As a simple example, consider the well-known Fisher Iris data depicted in Figure~\ref{fig:iris}.
In a model-based clustering scenario, the goal of estimation would be to learn the shapes of the three
four-dimensional distributions, one for each species, without the benefit of the species labels.  
Under the conditional independence assumption
of Equation~(\ref{condInd}), no bivariate plot should exhibit correlation within any of the categories.
Yet it is clear in Figure~\ref{fig:iris}(a) that nonzero within-species correlation exists, so any correct
classifier of the three species would necessarily violate the conditional independence assumption.

To remedy this shortcoming,
the algorithm we propose in this article 
combines existing work on nonparametric mixture models
with the ideas of independent components analysis (ICA).  The basic idea of ICA, 
as elucidated, for example, by \cite{hyvarinen2002independent},
is to find a linear transformation of a multivariate dataset under which its coordinates are
as close to independent as possible.  

We can see the result of applying our algorithm to the
iris dataset in Figure~\ref{fig:iris}(b), in which each mixture component is associated with its own
linear transformation.  The categorizations displayed in this figure are based on the highest probability
of each point among the three possible categories, and we observe that 7 of the 150 points
are incorrectly classified---recall that the mixture model parameter estimates are calculated without
taking labels into account---but the correlation structure evident in the left-hand plots has been
eliminated.  In the remainder of this article, we describe the algorithm, discuss the potential
issues of identifiability that arise, and illustrate the algorithm's performance on three datasets 
including one in which $n=10{,}000$ and $r=144$.
The result of our work is an algorithm we call the NSMM-ICA algorithm, which we implement
in the {\bf icamix} package for R \citep{r2015}, available at
\url{http://cran.r-project.org/web/packages/icamix/index.html}. 

The novelty of the current article is its combination of the non-parametric mixture structure with
the linear transformations of ICA;
previous work on model-based clustering using ICA has imposed parametric assumptions on the component density functions.
\cite{lee1998unsupervised} and \cite{lee2000ica} propose parametric ICA mixture models with algorithms based on the 
infomax principle for various unsupervised classification problems. 
\cite{shah2004unsupervised} apply the ICA mixture model methodology to the problem of unsupervised classification of 
hyperspectral or multispectral imagery where image data are captured at multiple or a continuous range of frequencies across the 
electromagnetic spectrum. 
This is an important application of remote sensing and land cover classification. 
\cite{palmer2008newton} 
derive an asymptotic Newton algorithm for Quasi-Maximum Likelihood estimation of the parametric ICA mixture model and presents its 
application to EEG segmentation. 
Finally, \cite{salazar2010} extend the ICA mixture model methodology of \cite{lee1998unsupervised}
and others by positing that the component density functions actually have the form of kernel density 
estimates---i.e., the components are themselves mixtures of parametric functions of a fixed
and known form (the kernel density itself), 
with one component per observation.  \cite{salazar2015} apply a similar idea in an
agglomerative clustering framework; however, in our work we assume that the number of 
mixture components is fixed and known.  

As mentioned above, our work drops the parametric assumption of all the previous ICA mixture model
literature that we are aware of.  
Yet it does bear some similarities to the work described in, for example,
\cite{salazar2010} and \cite{salazar2015}, because we also employ kernel density estimation 
in order to approximate the unknown underlying univariate density functions that we describe below.
There is the additional similarity that these papers minimize a Kullback-Leibler divergence, which
has a similar form to our penalized and smoothed Kullback-Leibler divergence objective function.  
However, a crucial distinction is that we do not assume the knowledge of any of the category labels 
{\em a priori}; our algorithm is designed to handle
completely unsupervised multivariate clustering problems.
In addition, there are methods other than ICA in the literature based on the same idea of exploring mutlivariate
data to determine coordinate systems having some desirable property such as inter-coordinate independence.  
Invariant co-ordinate selection \citep{tyler2009,miettinen2015}, or ICS, 
is one such method, and \cite{pena2010} in particular uses ICS to search for 
cluster structure in the data based on the eigenvalues of a kurtosis matrix.
This work does not assume an underlying non-parametric mixture 
structure as in the current article,
yet in principle it would be possible to combine non-parametric mixtures with ICS instead of ICA.

\section{The nonparametric ICA mixture model}
\label{sec:npICAmixmod}

Most previous work on model~(\ref{condInd}) assumes that we observe the random sample 
${\bf Y}_1, \ldots, {\bf Y}_n$.  However, in the current article we generalize this previous work
by adding the assumption that the observed data are ${\bf X}_1, \ldots, {\bf X}_n$, where
${\bf X}_i = \mathsf{A}_j {\bf Y}_i$ for some invertible $r\times r$ matrix $\mathsf{A}_j$, conditional
on ${\bf Y}_i$ being generated from the $j$th component density $q_j$.  In other words, we introduce additional
parameters $\mathsf{A}_1, \ldots, \mathsf{A}_m$, one for each mixture component,
consisting of the matrices that linearly transform the 
latent ${\bf Y}_i$ with independent coordinates into the observed ${\bf X_i}$.
When there is no mixture structure, this assumption is exactly the
independent component analysis (ICA) framework as described by \cite{hyvarinen2002independent}. 
NB:  The word ``component'' in ``ICA'' is replaced by ``coordinate'' or ``dimension'' in the terminology of this
article; here, ``component'' refers to one of the mixture densities.

To aid notation, let us define $q_{\mathsf{A}}$ for
any nonnegative function $q$ on $R^r$ and invertible $r\times r$ matrix $\mathsf{A}$ as
\begin{equation*}
q_{\mathsf{A}}({\bf x})=q(\mathsf{A}^{-1}{\bf x})|\det{\mathsf{A}}|^{-1},
\end{equation*}
which is the density function of a linearly transformed random variable having density $q$ 
after left-multiplication by $\mathsf{A}$.

Our ICA mixture model may thus be described formally as follows:  We observe a random sample
${\bf X}_1, \ldots, {\bf X}_n$ from the mixture density
\begin{equation}\label{ICAmixturedensity}
g({\bf x}) = 
\sum_{j=1}^m \lambda_j f_j({\bf x}),
\end{equation}
where 
\begin{equation}\label{ica1}
f_j({\bf x}) = ({q_j})_{\mathsf{A}_j}({\bf x})
\end{equation}
and
\begin{equation}\label{ICAdensity}
q_j({\bf y})=\prod\limits_{k=1}^{r}q_{jk}({y}_k).
\end{equation}
For each observed ${\bf X}_i$, we shall define the usual latent variables
\[
Z_{ij} = I\{ \mbox{ ${\bf X}_i$ is drawn from the $j$th mixture component} \}
\]
and ${\bf Y}_i = \mathsf{A}_j^{-1} {\bf X}_i$ for the unique $j$ such that $Z_{ij}=1$.
For estimation purposes, we write
\begin{equation*}
(e_j)_{\mathsf{A}_j} = \lambda_j f_j,
\end{equation*}
so $e_j({\bf x}) = \lambda_j q_j({\bf x})$. 
Since any constant multiple of ${\bf Y}_i$ can be absorbed into the $\mathsf{A}_j$ matrices, we
mitigate against non-identifiable parameters by further assuming
\begin{equation}\label{reduceAmbiguity}
\mbox{Var\,} Y_{ik}=1
\end{equation} 
for all $i$ and for $1\le k\le r$.  Finally, we assume for each $j$, $1\le j\le m$, 
at most one of the density functions $q_{j1}, \ldots, q_{jr}$ is a normal density function.
The reason for this last assumption is that ICA operates on standardized versions of the data,
and thus if a subset of the linearly transformed ${\bf Y}$ coordinates is multivariate normal, the standardized versions
are always standard multivariate normal so there is no way to uniquely identify an ICA transformation $\mathsf{A}_j$.
Thus, the non-normality assumption, along with 
assumptions (\ref{ica1}), (\ref{ICAdensity}), and (\ref{reduceAmbiguity}), 
are commonly used in the literature on ICA \citep{hyvarinen2002independent}.  
Since the $q_{jk}$ are not assumed to follow any parametric form,
we call the model 
\begin{equation}\label{NPICAmodel}
g({\bf x}) = 
\sum_{j=1}^m \lambda_j 
|\det \mathsf{A}_j |^{-1}
\prod\limits_{k=1}^{r} {q_{jk}}\left ( [\mathsf{A}_j^{-1}{\bf x} ] _k \right)
\end{equation}
a nonparametric ICA mixture model.

It is not known whether the parameters in Equation~(\ref{NPICAmodel}) are uniquely identifiable 
in the case of perfect information about the form of $g({\bf x})$.  Our empirical experience with our 
algorithm, of which we provide examples in Section~\ref{sec:applications}, 
suggests that parameter estimation is well-behaved, yet this important theoretical question 
remains.  For example, general identifiability does not follow directly from the facts that 
the $q_{jk}$ and $\lambda_j$ parameters are uniquely determined by Equation~(\ref{condInd})
and the $\mathsf{A}_j$ parameters are uniquely determined by Equation~(\ref{ICAdensity}) together
with ${\bf X}=\mathsf{A}_j {\bf Y}$, even under the usual assumptions that are stated above.
It may be possible to extend the methods of \cite{allman2009identifiability} to prove identifiability, but for now
we are in a situation analogous to the period just prior to the publication of that article, when 
estimation algorithms existed for cases in which only special cases of identifiability had been
addressed in the literature and no general identifiability result had yet been established.  
It is also important to realize that even in a case where parameters are theoretically within the
set of identifiable parameters, estimation may still be difficult when the true parameters happen to
be near the boundary of that set.  However, here again we can point to our empirical experience in an example
such as the iris dataset of Section~\ref{sec:introduction}.  Quite often, the iris species have been modeled 
in the literature as multivariate normal distributions, suggesting that the data are generated from
a set of parameters quite close to non-identifiability; yet in practice, our algorithm appears to
find reasonable $\mathsf{A}_j$ estimates as shown in Figure~\ref{fig:iris}(b).

\section{Parameter estimation}

This section introduces an MM-like algorithm that seeks to 
estimate the parameters ${\bf e}=(e_1,e_2,...,e_m)$ and 
$\mathsf{A}=(\mathsf{A}_1,\mathsf{A}_2,...,\mathsf{A}_m)$ by minimizing a function that
gives in some sense the distance between the empirical data distribution and the
theoretical mixture distribution determined by the parameters.

We begin by defining some operators that will aid notation.  Much of the development of this
section follows the recent work of \cite{levine2011maximum} and \cite{zhu2015}; the novelty here
is in the incorporation of the $\mathsf{A}_j$ matrices into the usual conditional independence
framework, which requires some delicacy.

First, we define the linear smoothing, or convolution, operators $S_h$ and $S_h^*$
on $L^1(R^r)$.
Let $s_h(\cdot,\cdot)\in L^1(R\times R)$ be a nonnegative kernel function satisfying
\begin{equation}\label{integrate_s}
\int s_h(v,z) \dif z=\int s_h(v,z) \dif v=1
\end{equation}
for $v,z \in R$.
Here, $h>0$ is a user-specified
tuning parameter often referred to as a bandwidth in smoothing contexts.  
For any $f\in L^1(R^r)$, define $S_hf$ and $S_h^*f$ by
\begin{equation*}
(S_hf)({\bf{x}})=\int\tilde{s}_h({\bf{x}},{\bf{u}})f({\bf{u}})\dif {\bf{u}} \quad\text{and}\quad (S_h^*f)({\bf{x}})=\int\tilde{s}_h({\bf{u}},{\bf{x}})f({\bf{u}})\dif {\bf{u}},
\end{equation*}
where
\begin{equation}\label{s_bar}
\tilde{s}_h({\bf{x}},{\bf{u}})=\prod\limits_{k=1}^{r}s_h(x_k,u_k) \qquad \text{for } {\bf{x}}, {\bf{u}} \in R^r.
\end{equation}
Furthermore, let
\begin{equation*}
(\mathcal{N}_hf)({\bf{x}})=
\exp[(S_h^*\log{f})({\bf{x}})].
\end{equation*}
Notice that $(\mathcal{N}_h)$ as an operator on $L^1(R^r)$ is nonlinear, 
as it is the exponentiation of the linear
operator $S_h$ applied to the logarithm of a function.  This nonlinear smoothing operator plays 
an important role in the algorithm.

Finally, we reproduce the projection-multiplication operator of \cite{zhu2015}, defined as 
\begin{equation}\label{definitionofP}
(Pf)({\bf{x}})=\ds\frac{\left[\prod\limits_{k=1}^{r} 
\displaystyle\int f({\bf{x}})\dif x_1\dif x_2\cdots\dif x_{k-1}\dif x_{k+1}
\cdots\dif x_r\right]}{\left[\int{f}\right]^{(r-1)}}.
\end{equation}
\cite{zhu2015} point out that
when $f$ is a density on $R^r$, the right side of (\ref{definitionofP}) simplifies because the denominator is 1,
and also that the $P$ and $S_h$ operators commute, i.e.,
 $(P \circ S_h)f=(S_h \circ P)f$.

Let us consider the hypothetical case of a known target density $g({\bf x})$, 
which we sometimes call the infinite sample size
case.  
To estimate the parameters ${\bf e}=(e_1,e_2,...,e_m)$ and 
$\mathsf{A}=(\mathsf{A}_1,\mathsf{A}_2,...,\mathsf{A}_m)$,
the idea is to minimize a measure of the distance
between $g$ and the mixture density determined by the parameters.  
Due to mathematical considerations explained in \cite{levine2011maximum},
we wish to first apply the nonlinear smoother to the mixture density and then
minimize the Kullback-Leibler distance between $g$ and this nonlinearly smoothed
density.
We therefore propose in this hypothetical case to minimize
\begin{equation*}
\int g({\bf x})\log\left[{g({\bf x})}/{\sum\limits_{j=1}^{m}[\mathcal{N}_he_j]_{\mathsf{A}_j}
({\bf x})}\right]\dif {\bf x}+\int\left[\sum\limits_{j=1}^{m}(e_j)_{\mathsf{A}_j}({\bf x})\right]\dif {\bf x}
\end{equation*}
with respect to ${\bf e}$ and $\mathsf{A}$.
To analyze an actual dataset, we would replace the $g$ density by the empirical distribution of the data, which leads
to the objective function
\begin{equation}\label{eq:OptimizationProblemICA}
\ell({\bf e},\mathsf{A})= - \sum_{i=1}^n \log \sum\limits_{j=1}^{m}[\mathcal{N}_he_j]_{\mathsf{A}_j}
({\bf x}_i) +\int\left[\sum\limits_{j=1}^{m}(e_j)_{\mathsf{A}_j}({\bf x})\right]\dif {\bf x}
\end{equation}
to be minimized with respect to ${\bf e}$ and $\mathsf{A}$.

Two aspects of Equation~(\ref{eq:OptimizationProblemICA}) are worth noticing.  First, the second integral is 
part of a penalty term whose presence guarantees a convenient property of the functional parameters
${\bf e}=(e_1,e_2,...,e_m)$ that minimize $\ell({\bf e},\mathsf{A})$
for a fixed $\mathsf{A}$, and this property is explained below
in Equation~(\ref{eq:sumisone}).  Second, the definition of $\ell({\bf e},\mathsf{A})$ uses
$\sum_j[\mathcal{N}_he_j] _{\mathsf{A}_j}({\bf x})$ instead of
$\sum_j\mathcal{N}_h[(e_j)_{\mathsf{A}_j}]({\bf x})$; that is, the nonlinear smoothing is applied
{\em before} the linear transformation.  The intuition is that after the transformation, the data are no longer
conditionally independent and standardized, so the smoothing would affect each dimension very 
differently if it were applied after the transformation.

%

An advantage of using the ${\bf e}$ parameters instead of ${\bf \lambda}$ and ${\bf q}$ 
is that the latter parameterization requires the constraint that every $q_j$ is a density function.  
With the ${\bf e}$ parameters, such a constraint is unnecessary:  
As a straightforward corollary of Theorem~2.1 of \cite{zhu2015}, 
any minimizer $(\tilde{\bf e}, \tilde{\mathsf{A}})$ of (\ref{eq:OptimizationProblemICA}) 
must satisfy
\begin{equation}\label{eq:sumisone}
\int\left[\sum\limits_{j=1}^{m}(e_j)_{\mathsf{A}_j}({\bf x})\right]\dif {\bf x}
=\int\sum_{j=1}^{m}\tilde e_j({\bf x})\dif {\bf x} =1.
\end{equation}

\section{The NSMM--ICA Algorithm}

Here, we derive an iterative algorithm for solving the main problem of minimizing Equation~(\ref{eq:OptimizationProblemICA}). 
The algorithm is based on the MM framework, which stands for majorization-minimization \cite{hunter2004tutorial} 
and which involves constructing
and minimizing an alternative to the $\ell({\bf e},\mathsf{A})$ function 
with respect to ${\bf e}$ and $\mathsf{A}$ at each iteration.

\subsection{Majorizing the objective function}
Given the current estimate ${\bf e}^{(0)}$ and $\mathsf{A}^{(0)}$, let us define
\begin{equation*}
w^{(0)}_j({\bf x})=\frac{\left[\mathcal{N}_he^{(0)}_j\right]_{\mathsf{A}^{(0)}_j}({\bf x})}
{\sum\limits_{j'=1}^{m}\left[\mathcal{N}_he^{(0)}_{j'}\right]_{\mathsf{A}^{(0)}_{j'}}({\bf x})}.
\end{equation*}
Since $\sum_j w^{(0)}_j({\bf x})=1$, Jensen's inequality gives
\begin{align}
&\ell({\bf e},\mathsf{A})-\ell({\bf e}^{(0)},\mathsf{A}^{(0)})\nonumber\\
&=-\int{g({\bf x})\log{\sum\limits_{j=1}^{m}
w^{(0)}_j({\bf x})
\frac{(\mathcal{N}_he_j)_{\mathsf{A}_j}({\bf x})}{(\mathcal{N}_he^{(0)}_j)_{\mathsf{A}^{(0)}_j}({\bf x})}}}\dif {\bf x}
+\int{\left(\sum\limits_{j=1}^{m}(e_j)_{\mathsf{A}_j}-\sum\limits_{j=1}^{m}(e^{(0)}_j)_{\mathsf{A}^{(0)}_j}\right)}\nonumber\\
&\leq
-\int{g({\bf x})\sum\limits_{j=1}^{m}
w^{(0)}_j({\bf x})
\log{\frac{(\mathcal{N}_he_j)_{\mathsf{A}_j}({\bf x})}
{(\mathcal{N}_he^{(0)}_j)_{\mathsf{A}^{(0)}_j}({\bf x})}}}\dif {\bf x}+\int{\left(\sum\limits_{j=1}^{m}
(e_j)_{\mathsf{A}_j}-\sum\limits_{j=1}^{m}(e^{(0)}_j)_{\mathsf{A}^{(0)}_j}\right)}. \nonumber
\end{align}
Thus, if we let
\begin{equation*}
b^{(0)}({\bf e},\mathsf{A})=-\int{g({\bf x})
\sum\limits_{j=1}^{m}w^{(0)}_j({\bf x})\cdot\log{{(\mathcal{N}_he_j)_{\mathsf{A}_j}({\bf x})}}}\dif {\bf x}
+\int{\left(\sum\limits_{j=1}^{m}(e_j)_{\mathsf{A}_j}\right)},
\end{equation*}
then
\begin{equation*}
\ell({\bf e},\mathsf{A})-\ell({\bf e}^{(0)},\mathsf{A}^{(0)})\leq b^{(0)}({\bf e},\mathsf{A})-b^{(0)}({\bf e}^{(0)},\mathsf{A}^{(0)}).
\end{equation*}
Therefore $b^{(0)}$ majorizes $\ell$ at $({\bf e}^{(0)}, \mathsf{A}^{(0)})$ up to an additive constant.  We conclude
that minimizing $b^{(0)}({\bf e},\mathsf{A})$ will create an MM algorithm, as explained by
\cite{hunter2004tutorial}, and taking the next estimate in the iterative algorithm to be the minimizer will
guarantee that the algorithm possesses a descent property.

\subsection{Minimizing the majorizer}\label{sec:alternating}
For each $j$, $1\leq j\leq m$, we wish to minimize
\begin{equation}\label{majorizedfunctionICA}
b_j^{(0)}(e_j,\mathsf{A}_j)=-\int{g({\bf x})w^{(0)}_j({\bf x})\cdot\log{{(\mathcal{N}_he_j)_{\mathsf{A}_j}({\bf x})}}}\dif {\bf x}
+\int{(e_j)_{\mathsf{A}_j}} ({\bf x}) \dif {\bf x}
\end{equation}
with respect to $e_j$ and $\mathsf{A}_j$.
Instead of finding a global minimizer for $b_j^{(0)}$, we first hold $\mathsf{A}_j$ fixed and minimize with respect to $e_j$,
then plug in the resulting update to $e_j$ and minimize with respect to $\mathsf{A}_j$.  
The resulting algorithm, which mimics the multiple ``conditional maximization'' steps of the
ECM algorithm \cite{meng1993}, does not actually minimize
$b_j^{(0)}$, but it does ensure that the next iteration achieves a smaller value of $b_j^{(0)}$.
This property is enough to guarantee the descent property, which states that the 
value of the objective function decreases at each iteration of the algorithm.

We find that Equation~(\ref{majorizedfunctionICA}) has a closed-form minimizer as a function of 
$e_j$ when $\mathsf{A}_j$ is held fixed.
\begin{proposition}\label{EMINIMIZER}
The minimizer of Equation~(\ref{majorizedfunctionICA}) with respect to 
$e_j$, with $\mathsf{A}_j$ held fixed, is
\begin{equation}\label{ICANSMMupdate}
\hat{e}_j({\bf u})=\frac{|\det{\mathsf{A}_j}|}{\left[\int{g(\mathsf{A}_j{\bf x})w^{(0)}_j(\mathsf{A}_j{\bf x})}\text{ d}{\bf x}\right]^{r-1}}
\cdot  \prod\limits_{k=1}^{r}\int{g(\mathsf{A}_j{\bf y})w^{(0)}_j(\mathsf{A}_j{\bf y})s_h(u_k,y_k)}\dif {\bf y}.
\end{equation}
\end{proposition}
A proof of Proposition~\ref{EMINIMIZER} is provided in Appendix~\ref{proof1}.

Equation~(\ref{ICANSMMupdate}) can be rewritten as
\begin{equation*}
\hat{e}_j({\bf u})=\left[P\circ S_h(|\det{\mathsf{A}_j}|\cdot (g\cdot w^{(0)}_j)\circ\mathsf{A}_j)\right]({\bf{u}})
\end{equation*}
using the $P$ operator of Equation~(\ref{definitionofP}).
In general, for any nonnegative function $f$ on $R^r$,
\begin{equation*}
S_h(f\circ\mathsf{A}_j)=(S_h)_{\mathsf{A}_j}(f)\circ\mathsf{A}_j,
\end{equation*}
where
\begin{equation*}
(S_h)_{\mathsf{A}_j}f({\bf{x}})=\int|\det{\mathsf{A}_j}|^{-1}\tilde{s}_h(\mathsf{A}_j^{-1}
{\bf{x}},\mathsf{A}_j^{-1}{\bf{u}})f({\bf{u}})\dif {\bf{u}}.
\end{equation*}
Thus, we may also write
\begin{align*}
(\hat{e}_j)_{\mathsf{A}_j}({\bf u})
&=\left[P_{\mathsf{A}_j}\circ(S_h)_{\mathsf{A}_j}(g\cdot w^{(0)}_j)\right](\bf u),
\end{align*}
where
\begin{equation*}
P_{\mathsf{A}_j}f({\bf u})=[P(f_{\mathsf{A}_j^{-1}})]_{\mathsf{A}_j}({\bf u})=[P(f\circ\mathsf{A}_j)](\mathsf{A}_j^{-1}{\bf{u}}).
\end{equation*}

Now let us turn to the minimization of Equation~(\ref{majorizedfunctionICA})
with respect to $\mathsf{A}_j$.
We first define
\begin{equation}\label{QikDependsOnAi}
\hat{q}_{jk}(u_k)=
\frac{|\det{\mathsf{A}_j}|}{\int{g(\mathsf{A}_j{\bf x})w^{(0)}_j(\mathsf{A}_j{\bf x})}\text{ d}{\bf x}}
\int{g(\mathsf{A}_j{\bf y})w^{(0)}_j(\mathsf{A}_j{\bf y})s_h(u_k,y_k)}\dif {\bf y}.
\end{equation}

If we apply the change of variables ${\bf x} = \mathsf{A}_j{\bf y}$ to Equation~(\ref{majorizedfunctionICA}) and
then plug in $\hat{e}_j({\bf u})$ into the resulting expression for $b_j^{(0)}(e_j,\mathsf{A}_j)$,
we find that minimizing the result with respect to $\mathsf{A}_j$ is equivalent to minimizing
\begin{equation}\label{linkToICA}
\log{|\det{\mathsf{A}_j}|}+\sum\limits_{k=1}^{r}\int \hat{q}_{jk}(u)\log \hat{q}_{jk} (u) \text{ d}u
\end{equation}
with respect to $\mathsf{A}_j$, where $\hat{q}_{jk}$ depends on $\mathsf{A}_j$ through (\ref{QikDependsOnAi}).

In Expression (\ref{linkToICA}), $\hat{q}_{jk}$ is the $k$th margin of the kernel smoothed 
version of $(g\cdot w^{(0)}_j)_{\mathsf{A}_j^{-1}}/\int{g\cdot w^{(0)}_j}$. 
In the discrete case where $dG({\bf x})$ is the empirical distribution, $\hat{q}_{jk}$ is the 
$k$th margin of the kernel density estimate based on the linearly transformed 
(by $\mathsf{A}_j^{-1}$) weighted observed data set, where the weight for the data point 
${\bf x}_i$ is $w^{(0)}_j({\bf x}_i)$. Let us denote this weighted data set by 
${\bf\mathsf{D}}^{(0)}_j$ and hence its linear transformation by 
$\mathsf{A}_i^{-1}{\bf\mathsf{D}}^{(0)}_j$. By (\ref{QikDependsOnAi}), 
the optimization mechanism at the current step views $\mathsf{A}_j^{-1}{\bf\mathsf{D}}^{(0)}_j$ 
as a weighted sample generated from the unknown density function $q_j$, where 
${\bf\mathsf{D}}^{(0)}_j$ 
is a weighted sample from the $j$th mixing component and $\mathsf{A}_j^{-1}$ is the 
matrix that recovers the associated ICA transformations. 
Let us call $\mathsf{A}_j^{-1}$ a recovering matrix. 

By Equation~(\ref{reduceAmbiguity}), we may treat
$|\det{A}_j|$ as fixed given the weighted data $\mathsf{A}_j^{-1}{\bf\mathsf{D}}^{(0)}_j$.
The second term in (\ref{linkToICA}) is an estimate of the sum of 
marginal entropies of $q_j$, which is equal, up to a term that does not involve $\mathsf{A}_j$, to the mutual 
information of marginals of $q_j$. According to \cite{hyvarinen2002independent}, minimizing mutual 
information in this setting---that is, minimizing the mutual information of $\mathsf{A}_j^{-1}{\bf S}$ given 
a randomly chosen weighted sample from ${\bf S}$)---can be acheived by existing ICA algorithms
such as the fastICA algorithm described in Section~\ref{sec:integratingNSMMwithFastICA}.

To summarize, the NSMM-ICA iterative algorithm will iterate as follows, where the parameters
at the $t$th iteration will be denoted by $({\bf e}^{(t)}, \mathsf{A}^{(t)})$:

{\bf Majorization Step: } For $1\leq j\leq m$, compute
\begin{equation*}
w^{(t)}_j({\bf x})=\frac{(\mathcal{N}_he^{(t)}_j)_{\mathsf{A}^{(t)}_j}({\bf x})}
{\sum\limits_{j=1}^{m}(\mathcal{N}_he^{(t)}_j)_{\mathsf{A}^{(t)}_j}({\bf x})}.
\end{equation*}

{\bf ICA Step: } Use the fastICA technique of
Section~\ref{sec:integratingNSMMwithFastICA}
to find $\mathsf{A}^{(t+1)}_j$ subject to (\ref{reduceAmbiguity})
that minimizes
\begin{equation*}
\sum\limits_{k=1}^{r}\int \hat{q}^{(t+1)}_{jk}(u) \log \hat{q}^{(t+1)}_{jk} (u) \text{ d}u,
\end{equation*}
for $j=1, \ldots, m$,
where $\hat{q}^{(t+1)}_{jk}(u_k)$ is defined in Equation~(\ref{QikDependsOnAi}).

{\bf Minimization Step: } Let
\begin{equation*}
e^{(t+1)}_j({\bf u})=\hat{\lambda}^{(t+1)}_j\hat{q}^{(t+1)}_j({\bf u})=
\hat{\lambda}^{(t+1)}_j\prod\limits_{k=1}^{r}\hat{q}^{(t+1)}_{jk}(u_k),
\end{equation*}
where
\begin{equation*}
\hat{\lambda}^{(t+1)}_j=\int{(g\cdot w^{(t)}_j)}.
\end{equation*}

\subsection{Practical Implementation of NSMM-ICA}\label{sec:integratingNSMMwithFastICA}
Section~\ref{sec:alternating} suggests alternating NSMM and ICA methods to form an iterative algorithm 
for the estimation of the nonparametric ICA mixture model.   This section describes the practical considerations
that went into the development of a package for R \cite{r2015}, called {\bf icamix}, that implements these ideas.

Empirical evidence suggests that NSMM and the npEM algorithm of \cite{benaglia2009like} tend to give very 
similar estimates \cite{levine2011maximum}. The reason is that usually $\mathcal{N}_hf$ is close to $f$ itself. 
This suggests that the smoothed version of the algorithm can reasonably be replaced by the non-smoothed version 
because the former is more computationally burdensome than the latter.   The decision to implement this non-smoothed
version affects only Step~1 of the algorithm below.  The result is an algorithm that fails to achieve the provable descent 
property of the smoothed version but which is much faster and which appears to result in nearly identical results for
most test problems.

Among the many ICA techniques available in the
literature, here we use the efficient and well-tested
FastICA of \cite{hyvarinen2002independent}.
At each iteration, FastICA will be applied to a weighted dataset, where the weight on observation $i$
for component $j$ is determined as the estimate, given the information available at the present iteration, 
of the probability that observation $i$ falls into component $j$. 

Assume we are given raw data as a matrix ${\mathsf{X}}^\top =\{\textbf{x}_1, \textbf{x}_2, ..., \textbf{x}_n\}^\top $, 
where $\textbf{x}_i=(x_{i1}, x_{i2}, x_{i3}, ..., x_{ir})^\top $ for $1\leq i\leq n$. 
We first choose a set of starting parameter values.  Since our algorithm, like any MM algorithm, 
finds at best a local minimum, it is possible that different starting values will lead to different solutions.
In the {\bf icamix} package, we begin with a $k$-means clustering solution, which assigns each data point 
to a distinct mixture component (and which itself has a stochastic element that allows for ease in choosing
multiple starting points).  Given this initial partition, straightforward estimation (e.g., using FastICA)
on the separate components leads to initial parameter values.
Then, our algorithm 
iterates through Steps~1 through 4 below until a convergence criterion is met.

{\bf Step 1.\ } For all $i$ and $j$, estimate the $j$th component weight for the $i$th observation using the non-smoothed densities:
\begin{equation*}
p^{(t)}_{ij}=\ds\frac{\lambda^{(t)}_{j}f_j^{(t)}({\bf x}_i)}{\sum\limits_{j\prime=1}^{m}\lambda^{(t)}_{j\prime}
f_{j\prime}^{(t)}({\bf x}_i)}
=\ds\frac{\lambda^{(t)}_{j}\left|\det{\mathsf{A}_j^{(t)}}\right|^{-1}\prod\limits_{k=1}^{r}q_{jk}^{(t)}
\left( \left[(\mathsf{A}_j^{(t)})^{-1}{\bf x}_i \right]_k\right)}{\sum\limits_{j\prime=1}^{m}\lambda^{(t)}_{j\prime}
\left|\det{\mathsf{A}_{j\prime}^{(t)}}\right|^{-1}\prod\limits_{k=1}^{r}q_{j\prime,k}^{(t)}
\left( \left[(\mathsf{A}_{j\prime}^{(t)})^{-1}{\bf x}_i \right]_k\right)}.
\end{equation*}

\noindent Steps 2, 3a, 3b, 3c, and 4 are now repeated separately for each value of $j$, $1\le j\le m$:

{\bf Step 2.\ } Update the ${\bf \lambda}$ parameters:
\begin{equation*}
\lambda^{(t+1)}_j=\ds\frac{1}{n}\sum\limits_{i=1}^{n}p^{(t)}_{ij}.
\end{equation*}

{\bf Step 3a.\ } Centering FastICA step for component $j$:  For each $i$, define
\begin{equation*}
\tilde{\bf x}_i \leftarrow {\bf x}_i-\ds\frac{\sum\limits_{i=1}^{n}{\bf x}_ip_{ij}^{(t)}}{\sum\limits_{i=1}^{n}p_{ij}^{(t)}}.
\end{equation*}

{\bf Step 3b.\ } Decorrelating FastICA step for component $j$: We first obtain the eigenvalue decomposition as
\begin{equation*}
\ds\frac{\sum\limits_{i=1}^{n}\tilde{\bf x}_i\tilde{\bf x}_i^\top p_{ij}^{(t)}}{\sum\limits_{i=1}^{n}p_{ij}^{(t)}}=E_jD_jE_j^\top ,
\end{equation*}
then let $V_j=E_jD_j^{-1/2}E_j^\top$ and ${\bf z}_{ij}=V_j\tilde{\bf x}_i$ for $i=1, \ldots, n$.  
Here, the notation ${\bf z}_{ij}$ refers to the $j$th component version of
the $i$th observation of the ${\bf z}$ vector, which is $r$-dimensional.
Therefore,
\begin{equation}\label{eq:standardized}
\ds\frac{\sum\limits_{i=1}^{n}{\bf z}_{ij}{\bf z}_{ij}^\top p_{ij}^{(t)}}{\sum\limits_{i=1}^{n}p_{ij}^{(t)}}=V_jE_jD_jE_j^\top V_j^\top =I.
\end{equation}
The transformed data ${\bf z}_{ij}$ with weights $p_{ij}^{(t)}$, $1\leq i\leq n$, thus have 
their coordinates uncorrelated and standardized according to (\ref{eq:standardized}). 
Since ${\bf Z}_j=V_j\tilde{\bf X}=V_j\mathsf{A}_j{\bf S}$, we need to first estimate $\left(V_j\mathsf{A}_j\right)^{-1}$, 
of which the $i$th row is the same as ${\bf w}_{ij}$ in (\ref{eq:updatewi}) below, and multiply it by 
$V_j$ on the right to get an update of $\mathsf{A}_j^{-1}$.

{\bf Step 3c.\ } Symmetric orthogonalization FastICA step for component $j$: 
At this step, we enter an internal loop that ultimately results in the update of the 
$\mathsf{A}_j$ matrix.  
Let $W_j=[{\bf w}_{1j}, {\bf w}_{2j}, ..., {\bf w}_{rj}]^\top $ for $r$-dimensional 
unit length vectors ${\bf w}_{1j}, {\bf w}_{2j}, ..., {\bf w}_{rj}$.  The first time we enter Step~3c, 
we may simply take 
${\bf w}_{ij}$ to be the $i$th standard basis vector for each $j$, and at succeeding
iterations we take these ${\bf w}_{ij}$ to be rows of the most recent $W_j$ matrix.

The inner loop then proceeds by updating the
${\bf w}_{ij}$ from their previous values according to
\begin{equation}\label{eq:updatewi}
{\bf w}_{ij}\leftarrow\ds\frac{\sum\limits_{i=1}^{n}{\bf z}_{ij}g({\bf w}_{ij}^\top {\bf z}_{ij})p_{ij}^{(t)}}
{\sum\limits_{i=1}^{n}p_{ij}^{(t)}}-{\bf w}_{ij}\ds
\frac{\sum\limits_{i=1}^{n}g\prime({\bf w}_{ij}^\top {\bf z}_{ij})p_{ij}^{(t)}}
{ \sum\limits_{i=1}^{n}p_{ij}^{(t)} },
\end{equation}
where $g$ may be chosen to be either $g(y)=\text{tanh}(\alpha_1y)$ for some 
$1\leq\alpha_1\leq2$ or $g(y)=y\exp(-y^2/2)$ \citep{hyvarinen2002independent}.
Let $W_j=[{\bf w}_{1j}, {\bf w}_{2j}, ..., {\bf w}_{rj}]^\top $ and then symmetrize and orthogonalize by
\begin{equation}\label{eq:symmetrizewi}
W_j\leftarrow(W_jW_j^\top )^{-1/2}W_j.
\end{equation}

Iteratively update the ${\bf w}_{ij}, i=1,...,r$ using Equation (\ref{eq:updatewi}) and (\ref{eq:symmetrizewi}) 
until convergence is achieved. More precisely, we choose a tolerance $\tau$ and stop updating when
\begin{equation*}
\max\limits_{1\leq i\leq r}\left\{\Bigl|\left({\bf w}_{ij}^{\rm (previous)}\right)^\top \cdot {\bf w}_{ij}^{\rm(current)}-1\Bigr|\right\}\leq \tau.
\end{equation*}

Finally, set
\begin{equation*}
\mathsf{A}_j^{(t+1)}=V_j^{-1}W_j^{-1}.
\end{equation*}

{\bf Step 4.\ } Non-parametric density estimation step: For all $j$ and $k$, let
\begin{equation*}
q^{(t+1)}_{jk}({\bf u})=
\left( \sum\limits_{i=1}^{n}p^{(t+1)}_{ij} \right) ^{-1}
\sum\limits_{i=1}^{n}p^{(t+1)}_{ij}\ds
\frac{1}{h}K\left(\ds\frac{{\bf u}- \left[(\mathsf{A}^{(t+1)}_j)^{-1}\textbf{x}_i \right]_k}{h}\right).
\end{equation*}

The R package {\bf icamix} makes use of the {\bf Rcpp} \cite{rcppart,rcppbook} and {\bf RcppArmadillo} \cite{rcpparmaart} 
packages for compiling and calling the core algorithms implemented in C++ code to speed up the calculations. 

In the discrete algorithm we have developed, a single fixed bandwidth calculated from the data will not be sensible, 
especially because the scale is now changing according to the ICA framework. Thus we propose an iterative scheme 
for choosing the bandwidth similar to that of \cite{benaglia2011bandwidth}, whereby
\begin{equation}\label{eq:ouriteratebandwidth}
h^{t+1}_{jk}=0.5\cdot\text{SD}^{t+1}_{jk}\cdot (n\lambda^{t+1}_{j})^{-0.2}=0.5\cdot (n\lambda^{t+1}_{j})^{-0.2}.
\end{equation}
For simplicity, we replace $\min{\{\text{SD}^{t+1}_{j,l}, \text{IQR}^{t+1}_{j,l}/1.349\}}$ in the original \cite{benaglia2011bandwidth}
formulation by $\text{SD}^{t+1}_{j,l}=1$.
We also propose the ad hoc coefficient of 0.5 rather than Silverman's 0.9 used by \cite{benaglia2011bandwidth}
in order to capture fine features of the density for better performance in the classification task. 
Our experience is that using 0.9 tends to oversmooth the estimated density. Simulation studies and applications 
we have run suggest that Equation~(\ref{eq:ouriteratebandwidth}) works well in practice.

When running the algorithm, we have determined that Step~4 dominates the computing time. 
By making use of Gaussian kernels and utilizing certain symmetric structure in evaluating some of Gaussians, we are able to 
lower the computing cost for the kernel density estimation step by about 50\% with respect to the {\bf mixtools} package. 
Further improvement might be possible via the Fast Gauss Transform \cite{raykar2005fast} and related techniques,
though we have not implemented these improvements.

\section{Applications}\label{sec:applications}
Here, we describe our experience applying the modified NSMM-ICA algorithm implemented in the {\bf icamix} 
package to several datasets of varying size.

\subsection{Italian Wine Classification}\label{sec:wine}

The Italian wine data set is another popular data set used for comparing various classifiers \cite{forina1988parvus,aeberhard1992comparison}. 
It contains results of a chemical analysis of wines grown in Italy but derived from three different cultivars. 
A total of 178 observations are recorded, each with 13 continuous attributes such as color intensity, magnesium and malic acid. 
There are 59, 71 and 48 instances in the first (Barolo), second (Grignolino) and third (Barbera) wine classes, respectively.

\begin{table}[ht]
\caption{Wine Data classifications by PCA+NSMM-ICA algorithm}
\centering
\begin{tabular}{l || c c c}
\hline\hline
\rule{0pt}{3.6ex}   & Class 1 & Class 2 & Class 3 \\ [1.5ex]
\hline
\rule{0pt}{3.6ex}  
Barolo & 59 & 0 & 0 \\
Grignolino & 6 & 61 & 4 \\
Barbera & 0 & 0 & 48 \\[1ex]
\hline
\end{tabular}
\label{table:winepcansmmicaclass}
\end{table}

\begin{figure}[htb]
    \centering
    \includegraphics[height=1.7in]{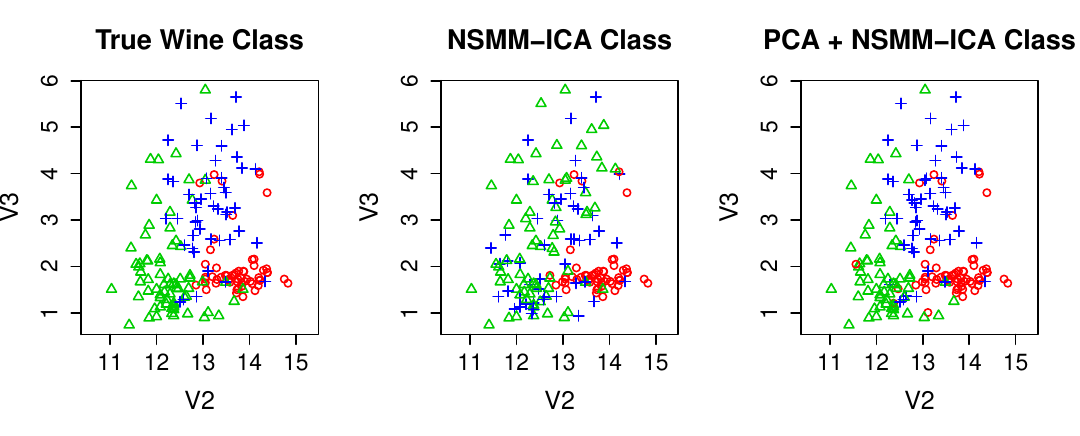}
    \caption{Wine Data: Comparison of true species Information (far left) and two results from our unsupervised learning algorithms.}
    \label{Ch4-figure:wineresults}
\end{figure}

If we feed the unlabeled data directly to the NSMM-ICA algorithm, we obtain a classification error rate equal to $28.65\%$, 
prompting us to consider remedies. it seems that given the small number of observations, the relatively large number of attributes 
may be somewhat challenging as there are too many parameters to estimate and some of the attributes may consist of noise. So 
instead of using all 13 attributes, we first run principal component analysis (PCA) on the attributes and then select the 5 PCA scores 
that explain the largest proportion of variance in the attributes. 
Finally, we run the NSMM-ICA algorithm on the data set with the chosen PCA scores as attributes. 
In this way, the classification performance improves quite a lot, giving a classification error rate equal to $5.62\%$. 
Hence, in situations with relatively large numbers of coordinates, it might be worthwhile to utilize a dimension reduction 
technique followed by the NSMM-ICA algorithm. Figure \ref{Ch4-figure:wineresults} shows a comparison of true species 
information and results from our unsupervised learning algorithms.

\subsection{Tone Data}\label{sec:tonedata}

The tone perception experiment and data were first introduced by \cite{cohen1984some}. These data have been analyzed by 
\cite{de1989mixtures}, \cite{viele2002modeling}, and \cite{hunter2012semiparametric} in the context of mixtures of regressions. 
In each trial of the experiment, a musician is presented with a fundamental tone plus a series of overtones determined by a 
stretching ratio. Then the musician is asked to tune an adjustable tone to one octave above the fundamental tone. Both the stretching 
ratio and the ratio of the adjusted tone to the fundamental are reported for each trial. There are five musicians involved in the experiment. 
However, the tone data set only contains 150 trials with the same musician. The problem of interest in conducting this experiment is to 
investigate the theory that the musician would either tune the tones to the nominal
octave at a ratio of 2:1 to the fundamental tone (i.e., the
interval memory hypothesis) or use the overtone to tune the tone to the stretching ratio (i.e., the partial matching hypothesis). 
The findings by \cite{hunter2012semiparametric} via modeling through a semi-parametric mixture of regressions conforms with the
latter theory.

\begin{figure}
   \centering
   \includegraphics[width=1\linewidth]{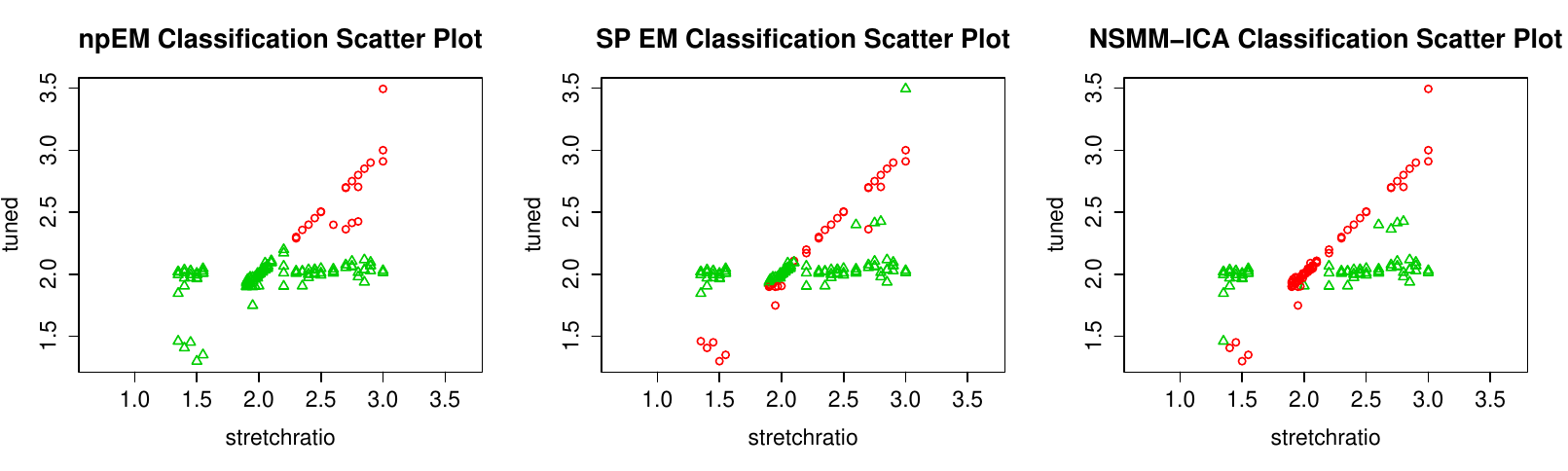}
   \caption{Comparison of three algorithms for fitting the tone dataset of \cite{cohen1984some}.  Only the plot labeled SP EM 
   explicitly assumes a mixture of regressions.  The npEM algorithm does not utilize ICA and therefore misses the
   two lines entirely.}
   \label{fig:tonecompare2}
 \end{figure}

For this unsupervised learning task, we run both the npEM algorithm by \cite{benaglia2009like} and our NSMM-ICA algorithm on the 
tone data. Figure \ref{fig:tonecompare2} shows that our NSMM-ICA algorithm does a good job of classification, very close to the
mixture-of-regressions results obtained by \cite{hunter2012semiparametric}, despite omitting any explicit assumption of 
regression structure. The reason why the results of the npEM algorithm shown in 
Figure~\ref{fig:tonecompare2} do not look nearly as good is because with regression lines that have nonzero slopes the mixture is 
far from being conditionally independent.  Thus, the additional ICA step in our algorithm is essential.

For comparing our result with that of \cite{hunter2012semiparametric},  we can use the estimated mixing weights obtained from 
NSMM-ICA to fit a weighted ordinary least squares model to obtain the regression coefficients. The results, summarized in 
Table~\ref{table:comparemixlsfits}, reflect the difference in estimated membership between SP EM and our NSMM-ICA 
primarily at the intersection of the two components, which is responsible for the noticeable difference in the estimated mixing weights.
 \begin{table}[ht]
\caption{Comparison of mixtures of least squares fits for the tone dataset of \cite{cohen1984some}.}
\centering
\begin{tabular}{c c || c c}
\hline\hline
\rule{0pt}{3.6ex} &  & SP EM & NSMM-ICA/Weighted LS \\ [1.5ex]
\hline
\rule{0pt}{3.6ex}  
component 1 & $\hat{\beta}_0$ & 1.77533 & 1.82215  \\
            & $\hat{\beta}_1$ & 0.11954 & 0.09076  \\
\hline
\rule{0pt}{3.6ex}
component 2 & $\hat{\beta}_0$ & 0.02121 & -0.12111 \\
            & $\hat{\beta}_1$ & 0.97929 & 1.05584  \\
\hline
\rule{0pt}{3.6ex}
$\hat\lambda_1$	&		          & 0.67653 & 0.46779  \\ [1ex]
\hline
\end{tabular}
\label{table:comparemixlsfits}
\end{table}

\subsection{Clustering images}\label{sec:learnimage}

Learning efficient codes for images obtained from different sources or contexts is an important problem in the area 
of image processing. The task involves extracting intrinsic structure in images by clustering and finding a complete set of efficient linear 
basis functions for each image source, which results in coefficient values being as statistically independent as possible. Techniques that 
utilize a parametric form of ICA mixture models have been proposed in \cite{lee1998unsupervised}, \cite{sejnowskiica}, and \cite{lee2000ica}.
Here, we apply the NSMM-ICA algorithm, which eliminates the parametric assumptions, 
to the task of unsupervised learning of image codes. 

\begin{figure}[htb]
    \centering
    \includegraphics[height=1.6in]{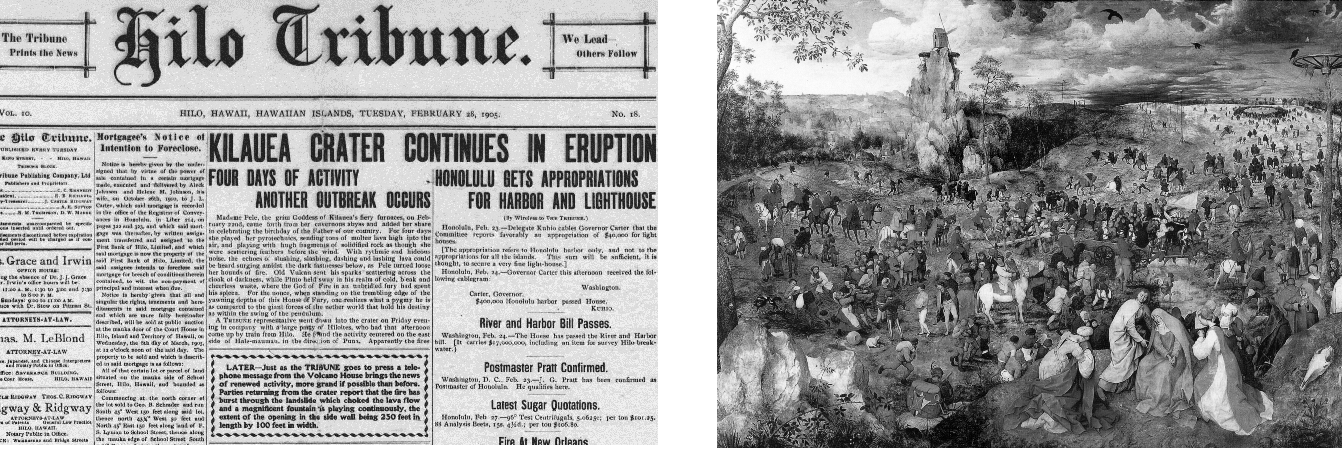}
    \caption{Two images as sources for data: Newspaper and painting.}
    \label{Ch4-figure:images}
\end{figure}

The two images shown in Figure~\ref{Ch4-figure:images} will be used as sources for the data set of the application. One is a painting 
image (2508$\times$ 1808 pixels) and the other is a newspaper image (2057$\times$1365 pixels). The images are transformed 
to grey scale:  Each pixel consists of a pixel intensity value ranging from 0 (black) to 1 (white).
We select 5000 12$\times$12 pixel patches randomly from each image. So the complete data set is of dimension 
10{,}000$\times$144. The NSMM-ICA algorithm converges after 19 iterations, which lasts a little less than 8 hours. 
Again each bandwidth is automatically learned by the iterative scheme we implemented. The result shows a very 
good recovery of the class-membership information, with a classification error rate of 1.2\%.

\begin{figure}[htb]
    \centering
    \includegraphics[width=.99\textwidth]{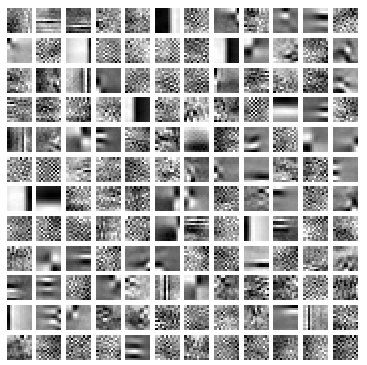}
    \caption{Learned basis functions for the newspaper image of
    Figure~\ref{Ch4-figure:images}. Each basis function, which is a 
    144-dimensional vector, is standardized to be within 0 and 1, then displayed as a $12\times12$ image patch.}
    \label{Ch4-figure:newspaper}
\end{figure}

\begin{figure}[htb]
    \centering
    \includegraphics[width=.99\textwidth]{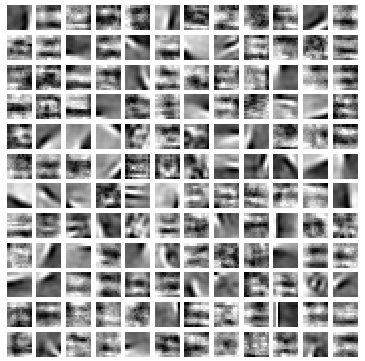}
    \caption{Learned basis functions for the painting image of
    Figure~\ref{Ch4-figure:images}. Each basis function, which is a 
    144-dimensional vector, is standardized to be within 0 and 1, then displayed as a $12\times12$ image patch.}
    \label{Ch4-figure:picture}
\end{figure}

The learned basis functions (i.e., a basis for the linear space of the pixel patches) show interesting patterns. 
Figures~\ref{Ch4-figure:newspaper} and~\ref{Ch4-figure:picture} show the basis functions for each image. The ones for the painting image appear smoother but 
more irregular, while the ones for the newspaper image look spottier and more regular.

\section{Discussion}\ 

Although the conditional independence model for multivariate data is promising due to the fact that its parameters are provably 
identifiable under weak conditions, even some simple datasets such as the well-known iris dataset clearly violate this 
conditional independence assumption.  The current article loosens this assumption, 
positing instead that each multivariate component can be linearly transformed to having 
independent coordinates.  This gives a much more flexible model-based clustering framework than the
conditional independence assumption alone.  Among the particular favorable features of
this extended model is that it allows for linear feature extraction, as illustrated 
by the tone data application of Section~\ref{sec:tonedata}.  

Yet much work remains to be done on this and similar methods.   Despite the favorable-looking results that we have obtained
on the datasets in this article, the theoretically important question of the identifiability of the parameters remains unresolved.
A related issue is the fact that ICA cannot operate in the setting of multivariate 
normal data, since in that case the standardized data are already standard 
multivariate normal, so that there is no
way to identify an ICA transformation.  Thus, with the increased flexibility of our current framework come additional
questions regarding conditions under which parameter identifiability holds.  

In addition, large-sample behavior of the 
NSMM estimator such as convergence rates is still not fully known;
however, recent work on alternative algorithms such as those of 
\cite{bonhomme2016a},
for which asymptotic properties have been proven, might provide 
suitable alternatives for estimation of the nonparametric mixture structure.
Similarly, alternatives to the ICA algorithm we employ here, such as ICS, are also possible, and it is possible that
exploiting the ability of ICS to search for interesting clustering features as in \cite{pena2010} could be exploited.  

In addition, there is the question of computational efficiency given the burden of estimating multiple univariate densities, particularly
for large datasets.
Our implementation of the algorithm currently replaces the smoothed NSMM 
portion of the algorithm by the non-smoothed npEM of \cite{benaglia2009like},
since empirical comparisons have suggested that these two algorithms often result in nearly the
same solutions.  
The {\bf icamix} package for R \citep{r2015}, available on CRAN, interweaves npEM with a weighted 
version of the FastICA algorithms \citep{hyvarinen2002independent}. The package also 
implements an automated and adaptive scheme for bandwidth selection that is based on the work of 
\cite{benaglia2011bandwidth}.   Further computing efficiencies may be attainable through the use of the 
Fast Gauss Transform.

It is important to remember that although this article compares some clustering solutions obtained via
our algorithm with those obtained using other techniques, such as k-means clustering, model-based clustering yields much more
than mere cluster memberships:  Not only does it assign each data point a probability vector of component membership,
but the statistical modeling of the individual components is often of great interest beyond the assignment of individuals to 
groups.  We did not explore the component density estimates obtained via our algorithm in this article, but
for an example that does so in the nonparametric mixture modeling literature, consider the
water-level dataset as analyzed in Section~5.2 of \cite{chauveau2015semi}.

All in all, given its flexibility and hence wide applicability, we believe that the novel approach to model-based
clustering presented here has the potential to be a useful alternative to existing approaches
based on parametric mixtures or mixtures that assume conditional independence.

\begin{appendix}

\section{Proof of Proposition~\ref{EMINIMIZER}} 
\label{proof1}

We assume that 
for each $1\leq j\leq m$, there exists $\theta_j>0$ such that
\begin{equation}\label{eq:CondInd}
e_j({\bf{x}})=\theta_j\prod\limits_{k=1}^{r}e_{jk}(x_k),
\end{equation}
where for each $k$, $1\leq k\leq r$, $e_{jk}\in L^1(R)$ is positive.  This overparameterization
is employed for the sake of convenience and
does not influence identifiability because we will never estimate $\theta_j$ separately.

The change of variables ${\bf x}=\mathsf{A}_j{\bf y}$ transforms Equation~(\ref{majorizedfunctionICA}) into
\begin{align}
b_j^{(0)}(e_j,\mathsf{A}_j) =  &
-\int{g(\mathsf{A}_j{\bf y})w^{(0)}_j(\mathsf{A}_j{\bf y})\cdot\log\left\{{(\mathcal{N}_he_j)({\bf y})
|\det{\mathsf{A}_j}|^{-1}}\right\}}|\det{\mathsf{A}_j}|\dif {\bf y} 
+\int{e_j}  ({\bf y}) \dif {\bf y}. 
\end{align}
Ignoring the term involving $\log |\det{\mathsf{A}_j}|^{-1}$ since it does not involve $e_j$, 
we find that minimizing $b_j^{(0)}(e_j,\mathsf{A}_j)$ as a function of $e_j$
with $\mathsf{A}_j$ fixed is equivalent to minimizing
\begin{equation}\label{eq:MinimizationICA}
-\int{g(\mathsf{A}_j{\bf y})w^{(0)}_j(\mathsf{A}_j{\bf y})\int{\bar{s}_h({\bf u},{\bf y})\log{e_j({\bf u})}}
|\det{\mathsf{A}_j}|\dif {\bf u}}\dif {\bf y}+\int{e_j}({\bf u}) \dif {\bf u},
\end{equation}
which by Equations~(\ref{integrate_s}), (\ref{s_bar}), and~(\ref{eq:CondInd}) equals
\begin{equation}\label{eq:optimizationwithfixedAi}
- \sum_{k=1}^r \iint{g(\mathsf{A}_j{\bf y})w^{(0)}_j(\mathsf{A}_j{\bf y}) s_h(u_k,y_k)\log{e_{jk}(u_k)}
|\det{\mathsf{A}_j}|\dif u_k }\dif {\bf y}
+\theta_j\int \prod_{k=1}^r{e_{jk}}(u_k) \dif u_k
\end{equation}
plus a term involving $\log \theta_j$ but none of the $e_{jk}$.

Picking a specific $k$ and
viewing Expression~(\ref{eq:optimizationwithfixedAi}) as an integral with respect to d$u_k$, we minimize it by minimizing 
the value of its integrand at each point. Differentiating with respect to $e_{jk}(u_k)$ and setting it equal to zero gives
\begin{equation*}
-\frac{\int{g(\mathsf{A}_j{\bf y})w^{(0)}_j(\mathsf{A}_j{\bf y})s_h(u_k,y_k)}|\det{\mathsf{A}_j}|\dif {\bf y}}{e_{jk}(u_k)}
+\theta_j\left[\prod\limits_{l\neq k}\int{e_{jk}(u_l)}\dif u_l\right]=0,
\end{equation*}
yielding
\begin{equation*}
\hat{e}_{jk}(u_k) \propto \int{g(\mathsf{A}_j{\bf y})w^{(0)}_j(\mathsf{A}_j{\bf y})s_h(u_k,y_k)}\dif {\bf y},
\end{equation*}
which implies
\begin{equation}\label{t0plugin20}
\hat{e}_{j}({\bf u}) =\alpha_j \prod\limits_{k=1}^{r}\int{g(\mathsf{A}_j{\bf y})w^{(0)}_j(\mathsf{A}_j{\bf y})s_h(u_k,y_k)}\dif {\bf y}
\end{equation}
for some constant $\alpha_j$.
To find $\alpha_j$, we plug (\ref{t0plugin20}) into (\ref{eq:MinimizationICA}), differentiate with respect to $\alpha_j$, and set the result equal to zero to obtain
\begin{equation*}
\alpha_j=\frac{|\det{\mathsf{A}_j}|}{\left[\int{g(\mathsf{A}_j{\bf x})w^{(0)}_j(\mathsf{A}_j{\bf x})}\text{ d}{\bf x}\right]^{r-1}},
\end{equation*}
which implies Equation~(\ref{ICANSMMupdate}).
\end{appendix}


\bibliographystyle{apalike}
\bibliography{Biblio-Database}   


\end{document}